\documentclass[%
 reprint,
groupedaddress,
 amsmath,amssymb,
 aps,
prb,
]{revtex4-1}

\usepackage{graphicx}
\usepackage{dcolumn}
\usepackage{bm}

\begin{document}

\newcommand{\be}{\begin{equation}}
\newcommand{\ee}[1]{\label{#1}\end{equation}}
\newcommand{\bem}{\begin{eqnarray}}
\newcommand{\eem}[1]{\label{#1}\end{eqnarray}}
\newcommand{\eq}[1]{Eq.~(\ref{#1})}
\newcommand{\Eq}[1]{Equation~(\ref{#1})}
\newcommand{\vp}[2]{[\mathbf{#1} \times \mathbf{#2}]}


\title{ Is spin superfluidity possible  in YIG films?}

\author{E.  B. Sonin}
 \affiliation{Racah Institute of Physics, Hebrew University of
Jerusalem, Givat Ram, Jerusalem 91904, Israel}

\date{\today}

\begin{abstract}
Recently it was suggested that stationary spin supercurrents (spin superfluidity) are possible in
the magnon condensate observed in yttrium-iron-garnet (YIG) magnetic films under strong external pumping.  Here we analyze this suggestion.  From topology of the equilibrium order parameter in YIG one must not expect energetic barriers making spin supercurrents metastable. However some small barriers of dynamical origin are possible  nevertheless. The critical phase gradient (analog of the Landau critical velocity in superfluids) is proportional to intensity of the coherent spin wave (number of condensed magnons). The conclusion is that although spin superfluidity in YIG films is possible in principle, the published claim of its observation is not justified.  

The analysis revealed that the widely accepted spin-wave spectrum in YIG films with magnetostatic and exchange interaction
required revision.   This led to revision of non-linear corrections, which determine stability of the magnon condensate with and without spin supercurrents.  
\end{abstract}

\maketitle


\section{Introduction}

Spin superfluidity  has already been discussed from 70s of the last century\cite{ES-78b,ES-82}. Its investigation continues nowadays (see recent reviews in Refs.~\onlinecite{Adv,Mac}).   The interest to this phenomenon was revived after emergence  of spintronics.  Manifestation of spin superfluidity is a stable spin supercurrent. Experimental observation of it in magnetically ordered solids would be an essential breakthrough in the condensed matter physics. A spin supercurrent is proportional to the  gradient of  the phase $\varphi$ (spin rotation angle in a plane) and is not accompanied by dissipation, in contrast to a dissipative spin diffusion current proportional to the gradient of spin density. 

In general the spin current proportional to the  gradient of  the phase $\varphi$ is ubiquitous and exists in any spin wave or domain wall, although in these cases variation of the phase $\varphi$ is small (very small in weak spin waves and not more then on the order  $\pi$ in domain walls and in disordered materials). Analogy with mass and charge persistent currents (supercurrents) arises when  at long (macroscopical) spatial intervals along streamlines the  phase variation is many times larger than $2\pi$. 
The supercurrent state is a helical spin structure, but in contrast to equilibrium  helical structures is metastable.

An elementary process of relaxation of the supercurrent is phase slip. In this process a vortex with $2\pi$ phase variation around it crosses streamlines of the supercurrent decreasing the total phase variation across streamlines by $2\pi$. Phase slips are suppressed by energetic barriers for vortex creation, which disappear when phase gradients reach critical values determined by the Landau criterion.

Recently \citet{Pokr} suggested (see also Ref.~\onlinecite{PokrD}) that spin superfluidity is possible in a coherent magnon condensate created in yttrium-iron-garnet (YIG) magnetic films by strong parametric pumping\cite{Dem6}, and   \citet{spinY} declared experimental detection of spin supercurrent in a decay of this condensate. Although the experimental evidence of spin superfluidity was challenged  \cite{YCom,LvovR} (see discussion in the end of the paper) the very  idea of spin superfluidity  in YIG films deserves a further analysis. In YIG the equilibrium order parameter in the spin space was not confined to some easy plane analogous to the order parameter complex plane in superfluids. The easy-plane  order parameter topology providing a barrier stabilizing a supercurrent was considered as a necessary condition for spin superfluidity.\cite{Adv,Mac} However, one cannot rule out that metastability of supercurrent states is provided by barriers not connected with topology of the equilibrium order parameter. The goal of the present paper was to investigate this possibility and to determine critical values of possible supercurrents at which metastability is lost.  

The critical supercurrents are determined from the principle similar to that of the Landau  criterion for superfluids: any weak perturbation of the current state  always increases  the energy, and therefore the current state is metastable.
This requires an analysis of nonlinear corrections to spin waves in the Landau--Lifshitz--Gilbert  (LLG) theory.  But calculation of nonlinear corrections is based on knowledge of the wave pattern and the wave spectrum in the linear theory. It was revealed that the commonly accepted and used up to now the linear theory of spin waves in YIG films\cite{Kal,Pokr,Tup,Rez} must be revised. This was done properly taking into account boundary conditions at film surfaces. For films now used in experiments on coherent magnon condensation the boundary problem in the presence of exchange and magnetostatic interaction has an accurate analytical solution, which gives  the wave pattern and the wave spectrum different from known before. This is important for the stability analysis  of the magnon condensate with and without spin supercurrents.  

Section \ref{TopCur} discusses connection of metastability of current states and topology of the order parameter space, which is a continuum of all degenerate ground states emerging from continuous symmetry (gauge symmetry in superfluids, rotational symmetry of the spin space in ferromagnets). It is also demonstrated how  non-equilibrium state of spin precession  supported by magnon pumping creates an effective ``easy plane'' for the order parameter, which allows metastable spin supercurrents.
Section \ref{LLT} reviews the LLG theory and the dispersion relation of linear plane spin waves in YIG bulk.   Section \ref{FT} considers linear spin waves in YIG film and determines their pattern and dispersion relation solving the boundary problem in the presence of the exchange  and the magnetostatic interaction. The result differs from known before, and the origin of this difference is discussed. Section~\ref{stB} analyzes nonlinear corrections, which determine stability of the coherent magnon condensate and the distribution of magnons between two energy minima in the $\bm k$ space. The quasi-equilibrium approach fixing the total number of magnons  does not predict stable condensate, in conflict with observation of the coherent condensate in experiments. It was suggested to modify the quasi-equilibrium  approach  fixing magnon numbers condensed in any of two minima,  but not only their total number. Finally Sec.~\ref{LC} derives  critical gradients in supercurrents from the Landau criterion generalized on spin superfluidity. The last section \ref{DC} discusses and compares the results of the present work with results of previous investigations. 

\section{Topology and  superfluid spin currents } \label{TopCur}

\begin{figure}[t]
\includegraphics[width=.5\textwidth]{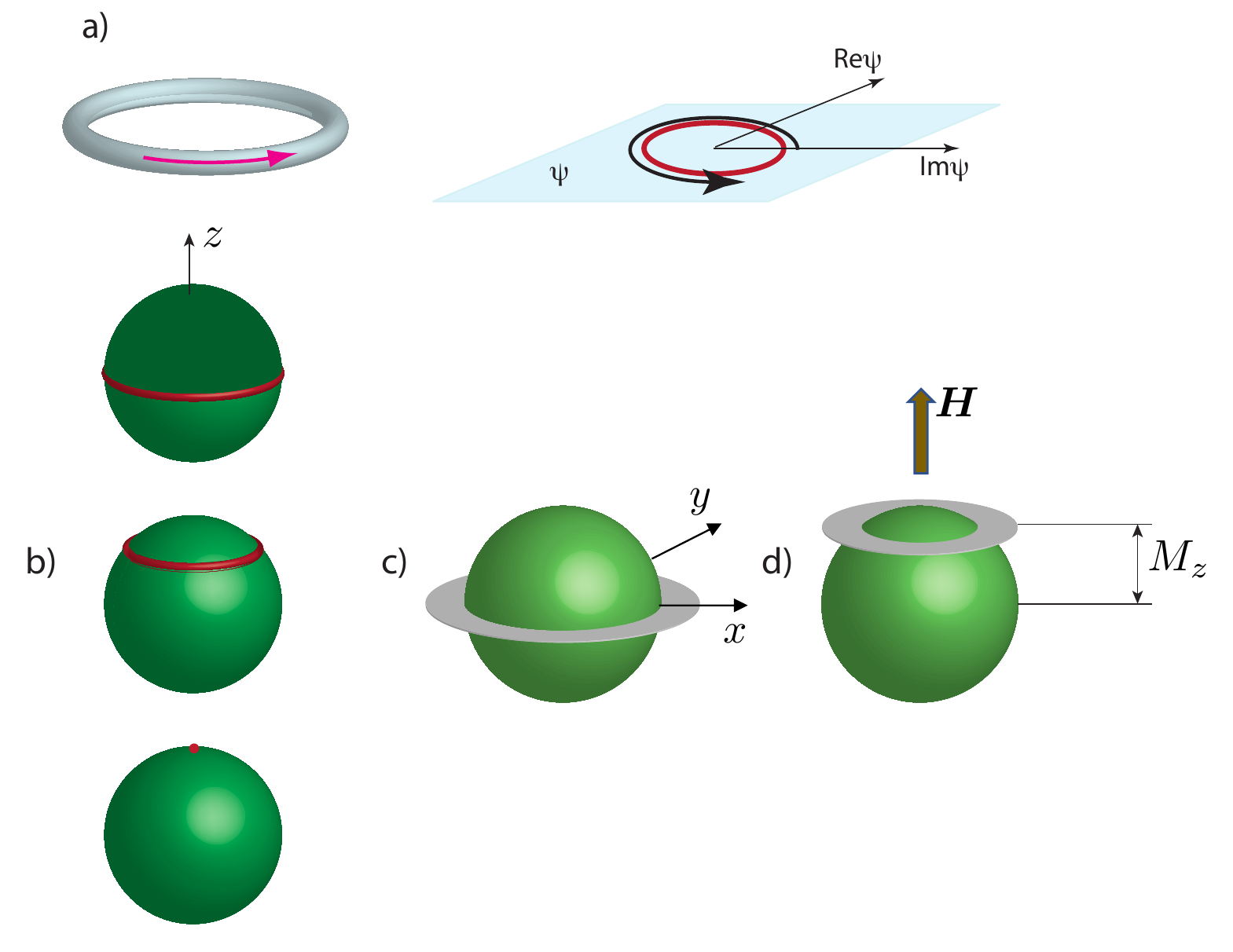}
\caption[]{Mapping of current states on the order parameter space.\newline
a) Mass currents in superfluids. The current state in torus maps on a circumference of radius $|\psi|$ on the complex plane $\psi$. \newline
b) Spin currents in an isotropic ferromagnet. The current state in torus maps on an equatorial circumference on the sphere of radius $M$ (top). Continuous shift of mapping on the sphere (middle) reduces it to a point at the northern pole (bottom), which corresponds to the ground state without currents. \newline
c) Spin currents in an easy-plane ferromagnet. Easy-plane anisotropy contracts the order parameter space to an equatorial circumference in the $xy$ plane topologically equivalent to the order parameter space in superfluids. \newline
d) Spin currents in an isotropic ferromagnet in a magnetic field parallel to the axis $z$ with nonequilibrium magnetization $M_z$ supported by magnon pumping. Spin is confined in  the plane parallel to the $xy$ plane but close to the northern pole. This plane is an ``easy plane'' of dynamical origin.  }
\label{Fig02}
\end{figure}

A knowledge on why superfluid currents can be metastable is provided by the analysis of topology of the order parameter space.
Spin superfluidity was suggested by the analogy with the more commonly  known mass superfluidity, and we start from discussion of the latter.  At the equilibrium the order parameter of a superfluid is a complex wave function   $\psi = \psi_0 e^{i\varphi}$, where the modulus $\psi_0$ of the wave function is a positive constant determined by minimization of the energy and the
 phase $\varphi$ is a degeneracy parameter since the energy does not depend on  $\varphi$ because of gauge invariance. Any from degenerate ground states in a closed annular channel (torus) maps on some point at a circumference $|\psi|=\psi_0$ in the complex plane $\psi$, while a
 current state with  the phase change $2\pi n$ around the torus maps onto a circumference (Fig.~\ref{Fig02}a) winding around the  circumference $n$ times. It is evident that it is impossible to change $n$ keeping the path on the circumference $|\psi|=\psi_0$ all the time. In the language of topology states with different $n$ belong to different classes, and
 $n$ is a {\em topological charge}. Only a phase slip can change it when the path in the complex plane leaves  the circumference. This should cost energy, which is spent on creation of a vortex crossing the cross-section of the torus channel and changing $n$ to $n-1$. The state with a vortex in the channel maps  on the full circle $|\psi| \leq \psi_0$.

If we consider transport of spin parallel to the axis $z$ the analog of the phase of the superfluid order parameter is the rotation  angle of the spin component in the plane $xy$, which we note also as $\varphi$. 
Here we neglect the processes, which break rotational invariance in spin space (analog of gauge invariance in. superfluids) and violate the conservation law for the total spin.  These processes can be of principle importance and were thoroughly investigated.\cite{Adv} But in the present discussion their effect can be ignored for the sake of simplicity.

In isotropic ferromagnets the order parameter space is a sphere of radius equal to the absolute value of the magnetization vector $\bm M$ (Fig.~\ref{Fig02}b). All points on this sphere correspond  to the same energy of the ground state.  Suppose we created the spin current state with monotonously varying  phase $\varphi$ in a torus. This state maps on the equatorial circumference on the order parameter sphere. Topology allows to continuously shift the circumference and to reduce it to the point of the northern pole.  During this process shown in Fig.~\ref{Fig02}b  the path remains on the sphere all the time and therefore no energetic barrier resists to the transformation. Thus metastability of the current state is not expected.

In a ferromagnet with easy-plane anisotropy the order parameter space contracts from the sphere to an equatorial circumference in the $xy$ plane. This makes the order parameter space topologically equivalent to that in superfluids (Fig.~\ref{Fig02}c).  Now transformation of the equatorial circumference to the point  shown in Fig.~\ref{Fig02}b costs anisotropy energy. This allows to expect metastable spin currents (supercurrents). They relax to the ground state via phase slips events, in which magnetic vortices cross spin current streamlines. States with vortices maps on  a hemisphere of radius $M$ either above or below the equator.

Up to now we considered states close to the equilibrium (ground) state. In a ferromagnet  in a magnetic field the equilibrium magnetization is parallel to the field. However,  by pumping magnons into the sample it is possible to tilt the magnetization with respect to the magnetic field. This creates a nonstationary state, in which the magnetization precesses around the magnetic field. Although the state is far from the true equilibrium, but it, nevertheless, is a state of minimal energy at fixed magnetization $M_z$. Because of inevitable spin relaxation  the state of uniform precession requires permanent pumping of spin and energy. However, if these processes violating the spin conservation law are weak, one can ignore them and treat the state as a quasi-equilibrium state. The state of uniform precession maps on a circumference parallel to the $xy$ plane, but in contrast to the easy-plane ferromagnet (Fig.~\ref{Fig02}c) the plane confining the precessing magnetization is much above the equator and not far the northern pole (Fig.~\ref{Fig02}d).  One can consider also a current  state, in which the phase (the rotation angle in the $xy$ plane) varies not only in time but also in space with a constant gradient. The current state will be metastable due to the same reason as in an easy-plane ferromagnet:  in order to relax via phase slips the magnetization should go away from the circumference on which the state of uniform precession maps, and this increases the energy. Then the plane, in which the magnetization precesses, can be considered as an effective ``easy plane'' originating not from the equilibrium order parameter topology but created dynamically. Further  the concept of dynamical easy plane will be applied to YIG magnetic films with some modifications. They take into account that the spin conservation law is not exact due to magnetostatic energy and the precession is not uniform  since  spin waves in YIG films have the energy minima at non-zero wave vectors. In contrast to  the equilibrium state, stability of  the dynamically supported non-equilibrium state even without current is not for granted and must be checked.

In our discussion of topology we assumed that phase gradients were small and ignored the gradient-dependent (kinetic) energy. At growing gradient and gradient-dependent energy, we reach the critical gradient at which barriers making the supercurrent stable vanish. For superfluids the critical gradient (critical velocity) is determined from the famous Landau criterion. The analogous criterion was also known for spin superfluidity in easy-plane anti- and ferromagnets.\cite{Adv} In the present paper we derive this criterion for possible spin supercurrents in  YIG magnetic films with the easy plane of dynamical origin.

\section{ Landau--Lifshitz--Gilbert theory and linear spin waves in YIG bulk} \label{LLT}

The coherent state of magnons is nothing else but a classical spin wave, and one can use the classical equations of   the LLG theory.
In  the LLG theory the absolute value of the magnetization vector $\bm M$ does not vary in space and time, and the classical LLG equations are reduced to two equations  for only two independent magnetization components:
 \be
\dot M_x=-\gamma M_z{\delta {\cal H} \over \delta M_y} ,~~\dot M_y =\gamma M_z {\delta {\cal H} \over \delta M_x},
       \ee{he}
where  $\gamma$ is the gyromagnetic ratio and $\delta {\cal H} / \delta M_x$ and $\delta {\cal H} / \delta M_y$ are functional derivatives of the hamiltonian ${\cal H}$. The third LLG equation for $\dot M_z$ is not independent and can be derived from two equations (\ref{he}).  Instead of two real functions $M_x$ and $M_y$ one can introduce one complex function $\psi =M_x+iM_y$. The equation  for $\psi$ directly follows from and fully equivalent to the LLG equations (\ref{he}). By analogy with the theory of superfluids they call it the Gross--Pitaevskii equation.

There is another form of the equations in the LLG theory especially convenient for the analysis of spin transport. Magnetization dynamics is described in the terms of two independent variables, $M_z$ and the angle $\varphi$ of the magnetization rotation around the $z$ axis:
\be
\dot M_z = - {\delta {\cal H}\over \delta \varphi} =   - {\partial  {\cal H}\over \partial \varphi} +\nabla_i{\partial  {\cal H}\over \partial \nabla_i \varphi}     = -\bm \nabla \cdot \bm j +T_z,
      \ee{bal} 
\be
\dot \varphi = {\delta {\cal H}\over \delta M_z}
   \ee{}
These are the Hamilton equations for the pair of conjugated canonical variables ``moment--angle'' analogous to the conjugated pair ``momentum--coordinate''. The first equation is the balance equation for magnetization along the axis $z$ proportional to the $z$ component of spin density, and we introduced the magnetization current $\bm j$ and the torque $T_z$:
\be
\bm j = {\partial  {\cal H}\over \partial \bm\nabla \varphi} ,~~T_z=  - {\partial  {\cal H}\over \partial \varphi}.
    \ee{curdef}

There was decades-long discussion of ambiguity in definition of the spin current.  Ambiguity emerges because the continuity equation for $M_z$ contains the torque $T_z$, which violates the spin conservation law. Indeed, one can add any vector $\bm b$ to the magnetization current $\bm j$ and compensate it by adding the divergence  $\bm \nabla \cdot \bm b$  to the torque $T_z$. This does not affect the final balance.  There were numerous attempts to find a proper definition of the spin current. It was argued\cite{Adv} that no definition is more proper than others. But some definition can be more convenient than others,  and the convenience criterion may vary from case to case. The choice of definition should not affect final physical results like the choice of gauge in electrodynamics.

YIG  is a  ferrimagnet with complicated magnetic structure consisting of numerous sublattices.\cite{YIG} However at  slow degrees of freedom relevant for our analysis one can treat it simply as an isotropic ferromagnet\cite{Melk} with the spontaneous magnetization $\bm M$ described by the hamiltonian\bem
{\cal H }=\int\left[-\bm H\cdot \bm M +D{\nabla_i  \bm M  \cdot \nabla_i  \bm M\over 2} \right]d\bm r
\nonumber \\
+ \int {\bm \nabla \cdot \bm M(\bm r) \bm \nabla \cdot \bm M(\bm r_1)\over 2|\bm r-\bm r_1| } d\bm r\,d\bm r_1 .
  \eem{}
Here the first term is the Zeeman energy in the magnetic field  $\bm H$, the second term  $\propto D$ is the inhomogeneous exchange energy, and  the last one is the magnetostatic (dipolar) energy.
Let us consider a spin wave in a YIG bulk propagating in the plane $xz$ in  a magnetic field $\bm H$ parallel to the axis $z$. In  a weak spin wave
\be
M_z \approx M-{M_\perp^2\over 2M},~~ \bm \nabla \cdot \bm M \approx \nabla_x M_x,
   \ee{}
where $M_\perp=\sqrt{M_x^2+M_y^2}$, and the linearized equations of motion (\ref{he}) are
\bem
\dot M_x=-\gamma H M_y + \gamma DM (\nabla_x^2 M_y+\nabla_z^2 M_y),
\nonumber \\
\dot M_y =\gamma H M_x - \gamma DM (\nabla_x^2 M_x+\nabla_z^2 M_x)
\nonumber \\
-  \gamma M\nabla_x \left(\int {\nabla_x M_x(\bm r_1) \over |\bm r-\bm r_1| }d\bm r_1 \right).
     \eem{em}
The equations look as integro-differential equations because of the magnetostatic term in the equation for $M_y$. But applying the  Laplace operator $\nabla^2$ to this equation makes this term purely differential. After exclusion of any of two component $M_x$ or $M_y$ one receives a differential equation of the 6th order.

For the plane wave with the frequency $\omega$ and the wave vector $\bm k(k_x,0,k_z)$ Eqs.~(\ref{em}) become
\bem
-i\omega M_x=-\gamma M_y (H + DM k^2),
\nonumber \\
-i\omega M_y =\gamma M_x \left(H + DM k^2 +{4\pi M k_x^2\over k^2}\right) .
       \eem{LL}
A solution of linear equations is an elliptically polarized  running spin wave,
\bem
M_x =m_0 \cos (\bm k \cdot \bm r+\omega t), 
\nonumber \\
 M_y = \sqrt{1+{4\pi M  k_x^2 \over (H +DM  k^2)k^2}}m_0 \sin(\bm k \cdot \bm r+\omega t),
    \eem{rw}
with the wave vector $\bm k(k_x,0,k_z)$ and the frequency 
\be
\omega(k)= \gamma \sqrt{(H + DM k^2)\left(H + DM k^2 +{4\pi M k_x^2\over k^2}\right)}.
   \ee{disp}
 The energy density in the spin wave mode   is 
\be
E =  { m_0^2\over 2M }\left(H  +DM k ^2 +{4\pi Mk_x^2\over k^2}\right)\approx {\omega (M-\langle M_z\rangle)\over \gamma},
       \ee{}
where $\langle M_z\rangle$ is the averaged magnetization. In quantum-mechanical description  the magnon density $n_m=E/\hbar \omega$ differs from the difference of $M-\langle M_z\rangle$  only by a constant factor. 

\begin{figure}[b]
\includegraphics[width=.5\textwidth]{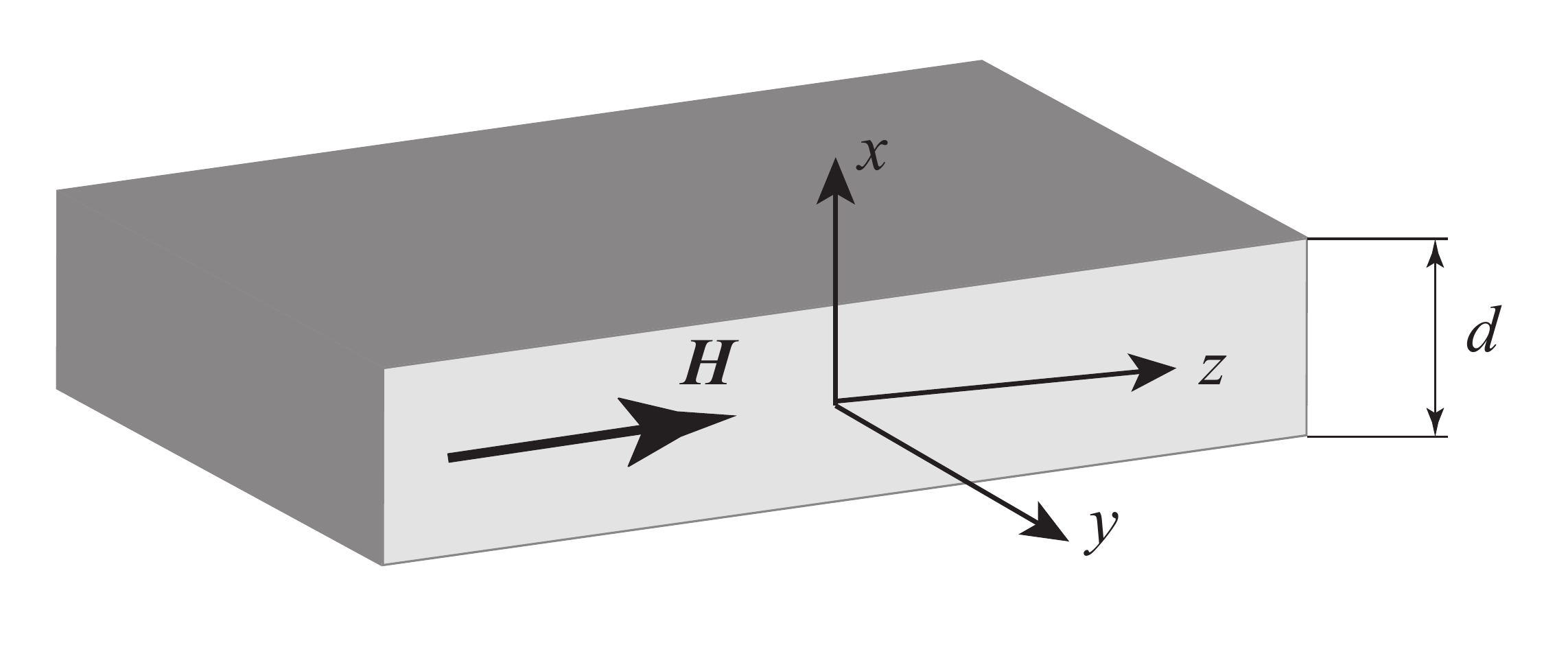}
\caption[]{The YIG film of thickness $d$ in a magnetic field $\bm H$ parallel to the axis $z$. }
\label{Fig1}
\end{figure}

\section{Spin waves in films, boundary conditions} \label{FT}

A spin wave propagating in the film  of thickness $d$ parallel to the plane $yz$ (Fig.~\ref{Fig1})  must satisfy the boundary conditions at two film surfaces $x=\pm d/2$. Neglecting the exchange interaction $\propto D$ the spin wave reduces to a magnetostatic wave investigated in the past by \citet{Dam}. The boundary conditions are imposed  on the magnetostatic magnetic field induced by magnetic charges $4\pi \bm \nabla \cdot \bm M(\bm r)$ and determined from the equation
\be
\bm \nabla \cdot (\bm h +4\pi \bm M)=0.
         \ee{}
The magnetostatic field is curl-free  and is given by
\be
\bm h =\bm \nabla \psi, ~~\psi(\bm r)=  \int { \bm \nabla \cdot \bm M(\bm r_1)\over |\bm r-\bm r_1| } d\bm r_1.
   \ee{mf}
At any film surface the tangential component of the magnetic field $\bm h$ and the normal component of the magnetic induction $\bm h +4\pi \bm M$ must be continuous.  For the magnetostatic mode $M_x \propto m_0\cos k_xx  e^{ik_zz -i\omega t}$ the magnetostatic potential inside the film is 
\be
\psi = -{4\pi M k_x \over k^2} \sin k_xx e^{ik_zz -i\omega t}.
     \ee{}
Outside the film at $x>d/2$  there is no magnetic charges and the magnetostatic potential must satisfy the Laplace equation $\Delta \psi=0$. 
Continuity  of the tangential component of the magnetic field $h_z=\nabla_z\psi$ at the film boundary requires continuity of  $\psi$, and  at $x>d/2$
\bem
\psi= -{4\pi M k_x \over k^2} \sin {k_xd\over 2} e^{k_z(d/2 -x)+ik_zz - i\omega t}.
    \eem{}
Continuity  of the  normal component of the magnetic induction $h_x +4\pi M_x=\nabla_x \psi +4\pi M_x$ also takes place if 
\be
\tan {k_xd\over 2} ={k_z\over k_x}.              
  \ee{kxd}
This equation  determines discrete values of $k_x$ for magnetostatic modes of \citet{Dam}.

In our case the exchange interaction cannot be ignored, and this imposes additional boundary conditions.  One cannot satisfy all boundary conditions by a single plane wave and must consider a superposition of plane waves with the same frequency $\omega$ and the wave number $k_z$ but with different values of $k_x$.

 The differential equations are of the 6th order in space. Correspondingly the dispersion relation   (\ref{disp}) at fixed $\omega$ and $k_z$ is a characteristic equation of the 6th order with respect to $k_x$ but is tri-quadratic (cubic with respect to $k_x^2$). The roots of the characteristic equation determine $k_x$ in the superposition. This approach was used in the past \cite{Wolf}. The first root of the cubic equation for $k_x^2$ yields a small real $k_x$, which determines the bulk mode with the frequency $\omega$. Other two roots  can be found analytically if the relevant wave number $k=\sqrt{k_z^2+k_x^2}$ is much smaller than $1/l_d$, where $l_d =\sqrt{D/\pi}$ is a small scale determined by the exchange energy.
 The values $k_{\pm}^2$ of two additional roots of the cubic equation for $k_x^2$  are negative and $k_{\pm}$ are imaginary and very large (on the order of $1/l_d$):
\bem
k_{\pm}^2  \approx {1\over D}\left( -2\pi-{H\over M} \pm \sqrt{4\pi^2+ {\omega^2\over \gamma ^2 M^2} }    \right) 
\nonumber \\
   \approx {1\over \pi l_d^2}\left( -2\pi-{H\over M} \pm \sqrt{4\pi^2+ {H^2\over  M^2} }    \right).
     \eem{}
These values correspond to evanescent modes  confined to surface layers of rather small width $l_d$.  

Close to the surface $x=d/2$ the boundary conditions are satisfied by a superposition of three modes:
\begin{widetext}
\bem   
M_x \propto \left[ \cos k_xx  +a_+ e ^{-p_+(d/2 -x) }+a_- e ^{-p_- (d/2 -x)}\right] e^{ik_zz -i\omega t},
\nonumber \\
M_y  \propto \left[\sqrt{1+{4\pi M  k_x^2 \over H+DMk^2}} \cos k_xx
  +a_+\sqrt{1+ {4\pi M   \over H -DM  p_+^2}} e ^{-p_+\left({d \over 2} -x\right)}+a_-\sqrt{1+ {4\pi M   \over H -DM  p_-^2}} e ^{-p_- \left({d \over 2} -x\right)}\right] e^{ik_zz -i\omega t},
  \nonumber \\
  \eem{}
\end{widetext}
where $p_\pm= ik_\pm$ are real and positive and $a_\pm$ are amplitudes of two evanescent modes.

The exchange boundary condition for unpinned spins \cite{Kal} are  $\nabla_x M_x=\nabla_x M_y=0$. They are satisfied if
\bem   
k_x  \sin {k_xd\over 2}  -a_+p_+ -a_- p_-=0,
\nonumber \\
k_x\sqrt{1+{4\pi M  k_x^2 \over H+DMk^2}}\sin {k_xd\over 2} 
\nonumber \\
 -a_+p_+\sqrt{1+ {4\pi M   \over H -DM  p_+^2}} 
 \nonumber \\
 -a_-p_-\sqrt{1+ {4\pi M   \over H -DM  p_-^2}}  =0.
  \eem{gr}
Repeating derivation of the magnetostatic boundary condition done above for magnetostatic modes one obtains
\bem
{k_x^2\over k ^2} \cos {k_xd\over 2}  +a_+ +a_-
\nonumber \\
={k_xk_z\over k^2} \sin {k_xd\over 2}  +{a_+k_z \over p_+}+{a_-k_z\over p_-}.
    \eem{msbc}
\Eq{gr} shows that the amplitudes of evanescent modes are of the order $a_\pm \sim  k_x\sin {k_xd\over 2} /p_\pm$. Then their contribution to the magnetostatic boundary  condition (\ref{msbc}) by a small factor $ k_z / p_\pm \sim k_zl_d$ less than the other terms and can be ignored. Eventually we return back to the equation  (\ref{kxd}) for $k_x$ obtained for magnetostatic waves  of \citet{Dam} without effects of exchange interaction.  Thus even though evanescent modes  are indispensable  for satisfying all boundary conditions they do not affect the shape of the wave in the most of the bulk.  

At large $k_z d$ \eq{kxd} yields $k_x=\pi/d$, and in the bulk  the magnetization components $M_x\sim \cos k_x x$ and $M_y \sim \cos k_x x$ vanish at the film surfaces. This automatically  satisfies the  exchange boundary conditions $M_x=M_y=0$ for pinned spins without adding evanescent modes.  Ignoring narrow surface layers where evanescent modes can be important, the plane wave propagating in the film plane is
\bem
M_x =\sqrt{2}m_0\cos {\pi x\over d} \cos (k_zz+\omega t), ~~
\nonumber \\
M_y = \sqrt{2}\left(1+{2\pi^3 M   \over H k_z^2d^2}\right)m_0\cos {\pi x\over d} \sin(k_zz+\omega t)
    \eem{pw}
independently from the exchange boundary conditions.    The wave frequency is 
\be
\omega(k_z) \approx  \gamma \left(H + DM k_z^2 +{2\pi^3 M \over k_z^2d^2}\right).
   \ee{disp1}
This dispersion relation differs from the spin-wave spectrum derived for YIG films by \citet{Kal} and widely used in the past, in particular, in articles addressing Bose--Einstein condensation and spin superfluidity  in YIG films \cite{Pokr,Tup,Rez}. 
  \citet{Kal}  received a dispersion relation, in which the term $2\pi M(1-e^{- k_zd}) / k_zd$ replaces the magnetostatic  contribution in our dispersion relation (\ref{disp1}) (the third term $\propto 1/k_z^2d^2$). Instead of solving differential equations \citet{Kal} approximately solved  the integro-differential equations. They approximated the magnetization distribution in space by a superposition of functions, which do not satisfy differential equations in the bulk. This is easily seen in the recent simplified derivation of their spectrum by \citet{Rez}. Rezende  approximated a spin wave in the film bulk by a superposition of plane-wave  modes with different values of $k_x$ as in our solution (our axis $x$ correspond to the axis $y$ of Rezende and vice versa). But Rezende's $k_x$ were not roots of the characteristic equation of the relevant system of differential equations. As a result, frequencies of modes in his superposition differ one from another and from the frequency given 
by the dispersion relation.  In particular, two of his modes have values $k_x =\pm ik_z$, for which $k^2=k_x^2+k_z^2$ vanishes and the spectrum (\ref{disp}) of a single plane spin wave gives an infinite frequency! Thus Rezende's superposition does not describe a proper monochromatic eigenmode at all. Correspondingly  the spectrum of Kalinikos and Slavin following from this superposition is invalid. 
  
   \begin{figure}[b]
\includegraphics[width=.5\textwidth]{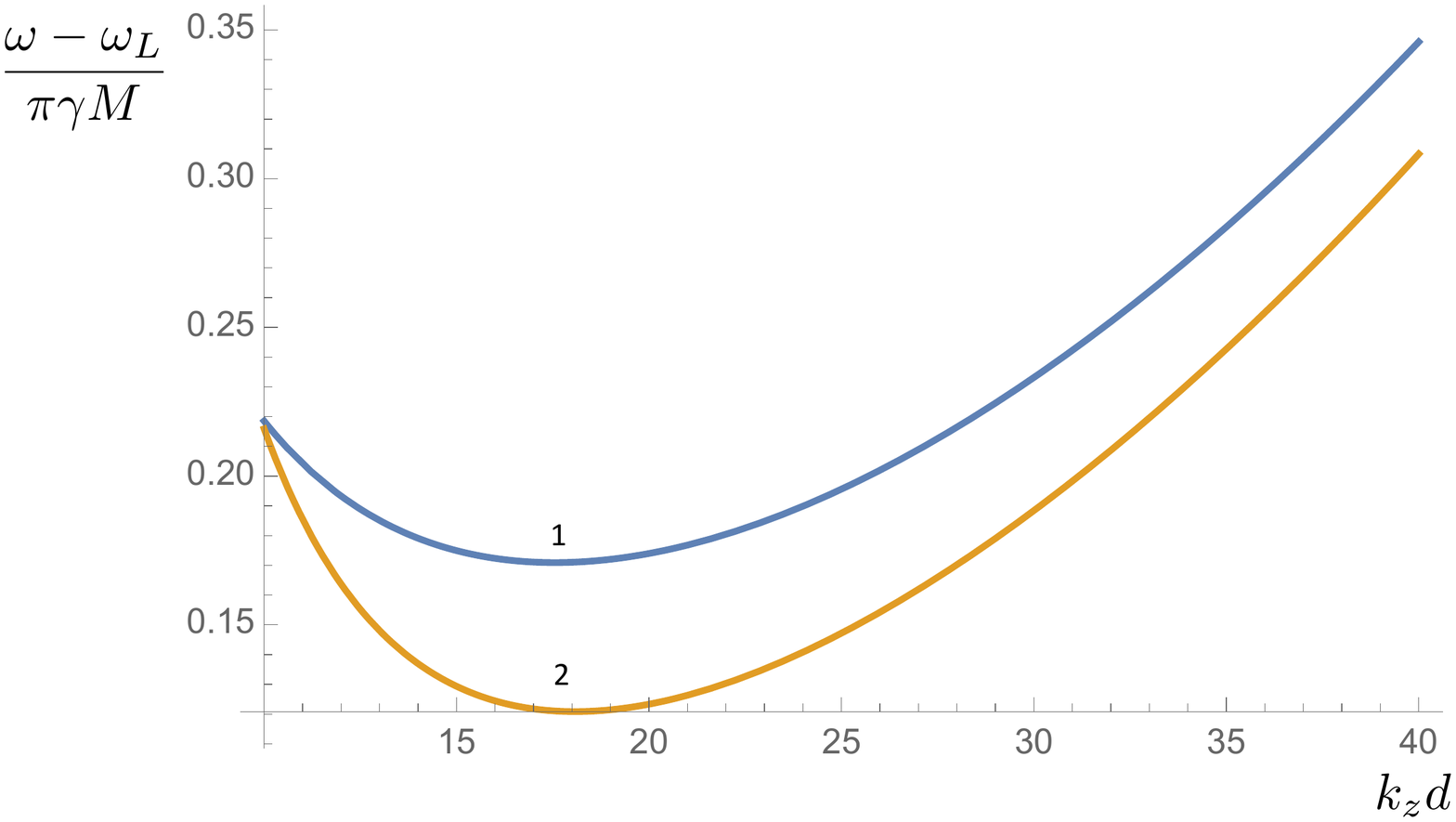}
\caption[]{Comparison of the linear spin-wave spectrum in a YIG film calculated by \citet{Kal} (curve 1) and in the present paper (curve 2). Here $\omega_L=\gamma H$ is the Larmor frequency. }
\label{Fig2a}
\end{figure}
  
The spectrum of Kalinikos and Slavin and the spectrum \eq{disp1} are compared in Fig.~\ref{Fig2a}. Quantitate difference between two spectra is not so dramatic. More important is that our analysis predicts an essentially different distribution of magnetization across the film. The component $M_x$ normal to the film approaches to zero close to the film surface (but still outside narrow boundary layers, where evanescent modes are important). On the other hand, according to   \citet{Rez}, in the approximation of \citet{Kal} variation of $M_x$ across the film is negligible.  This is important for evaluation of nonlinear corrections, which determine stability of supercurrent states investigated further in the paper. 
  
  Inaccuracy of the theory  of  \citet{Kal} has already been noticed by \citet{Kreis}. They calculated numerically the linear spin-wave spectrum in the microscopic theory and revealed that the numerically calculated spectrum lies lower than the spectrum of Kalinikos and Slavin as curve 2  in Fig.~\ref{Fig2a} calculated in the LLG theory. Agreement between the microscopic and macroscopic LLG theory  is not surprizing since all scales relevant for our analysis are larger than atomic.

\section{Coherent magnon condensate and its stability}\label{stB}

By strong parametric pumping \citet{Dem6} were able to create a coherent state of magnons condensed at states with lowest energies with non-zero wave vectors, which was called a magnon Bose--Einstein condensate. A condition for emerging of the magnon  condensate is that magnon-magnon interactions violating the spin conservation law are much weaker than interactions thermalizing the magnon gas. Despite the magnon gas required at least weak pumping for compensation of lost spin (magnons) it was treated as a quasi-equilibrium gas with fixed total number of magnons (see below).

\begin{figure}[t]
\includegraphics[width=.5\textwidth]{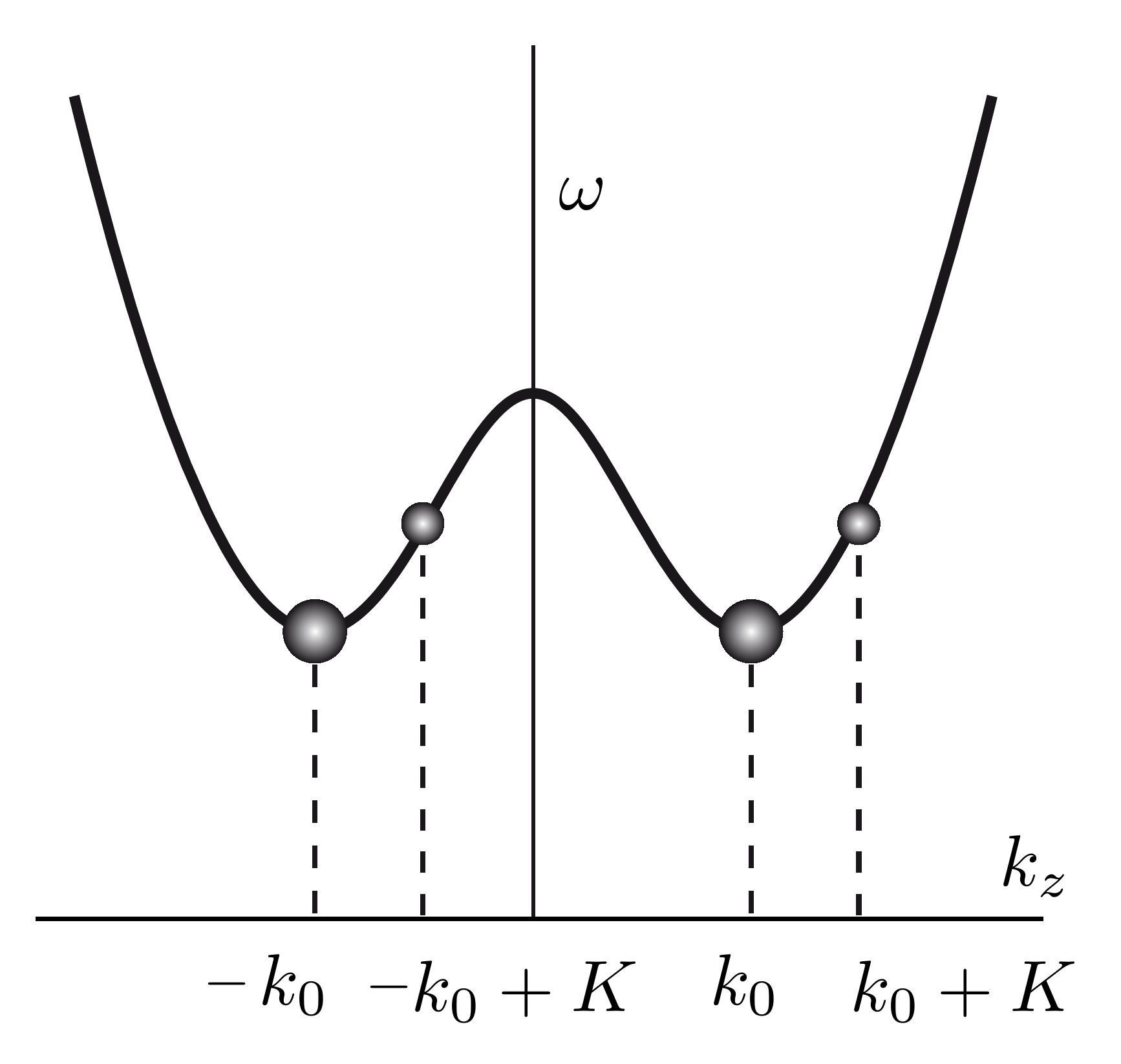}
\caption[]{The spin-wave spectrum in a YIG film. In the ground state the magnon condensate occupies two minima in the $k$ space with $k_z=\pm k_0$ (large circles). In the current state two parts of the condensate are shifted to $k = \pm k_0 +K$ (small circles). }
\label{Fig2}
\end{figure}

The energy and the frequency $\omega(k_z)$ given by \eq{disp1} have two degenerate minima\cite{Melk}  at finite $k_z =\pm k_0$ where magnons can condense  (Fig.~\ref{Fig2}). Here
\be
k_0=\left({2\pi^3\over Dd^2}\right)^{1/4}=\left({2\pi^2\over l_d^2 d^2}\right)^{1/4}.
     \ee{}
In the linear theory  the distribution of magnons between two condensates is arbitrary and does not affect the total energy (at fixed magnetization $\langle M_z\rangle$, i.e., at fixed condensate magnon density).  But  non-linear corrections lift this degeneracy.\cite{PokrPD} Let us consider the effect of  a non-linear  term $\propto M_\perp^4$ in the expansion for $\langle M_z\rangle$:
\be
\langle M_z \rangle =M-{\langle M_\perp^2 \rangle\over 2M}-{\langle M^4_\perp \rangle\over 8M^3}.
      \ee{}
The energy density of the condensate spin wave as a function of $M-\langle M_z \rangle$ is
\bem
E =H(M-\langle M_z \rangle)
\nonumber \\
+\left(DMk_z^2+{2\pi^3 M \over k_z^2d^2}\right)\left(M-\langle M_z\rangle-{\langle M^4_\perp \rangle\over 8M^3}\right). 
   \eem{}
For the running wave given by \eq{pw} (all magnons condensate in one minimum)  $\langle M^4_\perp \rangle =6(M-\langle M_z \rangle)^2$.  The sign of the nonlinear correction is negative. This corresponds to attraction between magnons, and the condensate is unstable. For the running wave (\ref{pw}) all other nonlinear corrections are smaller and cannot affect this conclusion.

However, if magnons condense in two minima there is another nonlinear term arising from the magnetostatic energy:
\be
E_{ms}  =\int {\nabla _z M_z(\bm r)  \nabla_z M_z(\bm r_1))\over 2|\bm r-\bm r_1| } d\bm r\,d\bm r_1.
    \ee{}
For the running wave this term is negligible compared to the term considered above, because $z$ variation of $M_z$ is weak.
But the nonlinear magnetostatic term  is maximal for the standing wave (two energy minima are equally populated by magnons): 
\bem
M_x =2m_0\cos {\pi x\over d} \cos k_zz\cos \omega t, 
\nonumber \\
 M_y = 2\left(1+{2\pi^3 M\over Hk_z^2d^2}\right)m_0\cos {\pi x\over d}\cos k_zz \sin\omega t.
    \eem{sw}
In the standing wave
\be
M_z =M- {m_0 ^2\over M} (1+\cos 2k_zz)  (1+\cos 2k_xx) ,
    \ee{}
and
\be
E_{ms}  ={3 \pi m_0^4 \over 8M^2} ={3 \pi (M-\langle M_z\rangle)^2 \over 2}.
    \ee{mse}
Now magnon interaction is repulsive. But this does not mean that  the standing-wave condensate is absolutely stable, because the interaction  energy  at fixed $\langle M_z \rangle$ decreases when the distribution of magnons between two condensates becomes more and more asymmetric. Eventually the condensate spin wave transforms to the running wave in which magnon-magnon interaction is attractive and the interaction energy is negative. Thus the magnon condensate cannot be stable! Then  inevitably  a question arises why a relatively stable long-living magnon condensate was observed. Instability of the magnon condensate in YIG films was already revealed earlier by \citet{Tup}. In order to explain the paradox that the magnon condensate was observed despite its expected instability, they referred to size effects. Another scenario is also possible. Apparently the quasi-equilibrium approach determining distribution of magnons between two energy minima from the condition of the minimal energy at fixed $\langle M_z \rangle$, i.e., at fixed total magnon number,  is not satisfactory, and   instead the magnon distribution between two minima must be received from  the dynamical balance taking into account spin pumping and spin relaxation. There is no evident reason why pumped magnons prefer to condensate in one minimum rather than in another, and \citet{Kopietz} numerically investigated  the dynamical process of the magnon condensate formation in the  LLG theory assuming that the two minima are filled symmetrically. \citet{Mal} solved numerically the Gross--Pitaevskii equation with added spin pumping and relaxation and found that sometimes asymmetric magnon distributions emerge, but only at asymmetric boundary conditions. Experimentally \citet{Dem12} revealed spatial periodic oscillations of magnon density, which are possible only if magnons condense in the  both energy minima.  

Apparently possible asymmetry of magnon distribution in the process of formation of the magnon condensate still deserves further investigations similar to those in Refs.~\onlinecite{Kopietz} and \onlinecite{Mal}, but it is beyond the scope of this work. Studying stability of current states (the next section) we shall use a modified quasi-equilibrium approach assuming that dynamical processes (spin pumping and relaxation) fix not only the total number of magnons but also  distribution of them between two energy minima.  We shall focus on a pure standing wave with symmetric magnon distribution in the $\bm k$ space for  which critical gradients are higher than for asymmetric distribution. Thus we look for the upper bound for critical gradients.

\section{Spin-supercurrent state and its stability (Landau criterion)} \label{LC}

The phase variation in space in the magnon condensate depends on distribution of magnons between two energy minima.
In the running wave (\ref{pw})
\be
\varphi = \arctan{M_y\over M_x}=\omega t + k_zz +{\pi^3 M \over k_z^2d^2} \sin 2(\omega t+k_zz), 
    \ee{}
while in the standing wave  
 \be
\varphi =\omega t +{\pi^3 M \over k_z^2d^2} \sin 2\omega t. 
    \ee{}

Thus apart from nonessential small periodical oscillations the phase gradient is $\nabla_z \varphi =k_z$ in the running wave but vanishes in the standing wave. 

In the standing wave the magnetization (spin) current appears if the wave numbers $k_z$  of two condensates differ from $\pm k_0$ (Fig.~\ref{Fig2}), and neglecting weak ellipticity \begin{widetext}
\bem
M_x =m_0 \cos k_xx[ \cos (k_0z+Kz+ \omega t)+\cos(k_0z-Kz-\omega t)]
=2m_0 \cos k_xx \cos k_0z\cos (Kz+ \omega t),
\nonumber \\
 M_y =m_0\cos k_xx[ \sin (k_0z+Kz+\omega t)-\sin(k_0z-Kz-\omega t)]
 =2m_0 \cos k_xx \cos k_0z\sin (Kz+ \omega t).
    \eem{SW}
\end{widetext}

Thus  $\nabla_z\varphi=K=k_z-k_0 \ll k_0$. Keeping the magnetization $\langle M_z \rangle$  fixed as before and taking into account the nonlinear magnetostatic term (\ref{mse})  the  energy in the spin-current state apart from some constant terms  is
 \be
 \Delta E = {d^2\omega (k_0)\over dk_z^2} {M-\langle M_z\rangle\over \gamma}{(\nabla_z\varphi)^2\over 2}+{3 \pi (M-\langle M_z\rangle)^2 \over 2},
        \ee{hMz}
where
\be
{d^2\omega (k_0)\over dk_z^2} =\gamma M\left(2D+{12\pi ^3\over k_0^4 d^2}\right)={16\pi ^3\gamma M\over k_0^4 d^2}.
  \ee{}

Stability of the spin-current state can be checked following the principal idea of the Landau criterion of superfluidity \cite{Adv}: If weak perturbations of the current state (creation of a quasiparticle in the Landau case) always increase  energy, the current state is  metastable. If there are perturbations decreasing the energy superfluid transport with suppressed dissipation is impossible. 
Let us consider slowly varying in space weak perturbations $m_z=M_z-\langle M_z \rangle $ and $\nabla_z\varphi'=\nabla_z \varphi-K $. Quadratic  in $m_z$ and $\nabla_z \varphi'$ terms in expansion of the energy (\ref{hMz})  are 
\bem
 \Delta E' = {d^2\omega (k_0)\over dk_z^2} \left[{M-\langle M_z \rangle\over \gamma}{(\nabla_z\varphi')^2\over 2}
\right.  \nonumber \\ \left.
-K {m_z\over \gamma}\nabla_z\varphi ' \right] +{3 \pi m_z^2 \over 2}.
       \eem{hMz0}
For stability of the supercurrent the quadratic form in perturbations   $m_z$ and $\nabla_z \varphi'$ must be always positive. This takes place as far as $\nabla_z\varphi=K$ is less than the critical value
\be
(\nabla_z\varphi)_{cr} =\sqrt{3\pi \gamma (M-\langle M_z \rangle)\over {d^2\omega (k_0)\over dk_z^2} }=  \sqrt{{ 3(M-\langle M_z \rangle)\over M}}{ k_0^2 d\over 4\pi }.
     \ee{}
This corresponds to the critical group magnon velocity
\be
v_{cr}={d^2\omega (k_0)\over dk_z^2} (\nabla_z\varphi)_{cr} ={4\pi^2 \gamma M\over k_0^2d} \sqrt{ 3(M-\langle M_z \rangle)\over M}.
     \ee{cr}
Note that applying our course of derivation to superfluid hydrodynamics one obtains exactly the Landau critical velocity equal to the sound velocity (see Sec.~2.1 in Ref.~\onlinecite{Adv}).

We conclude this section by estimation of the magnetization supercurrent $\bm j $   using  the canonical expression (\ref{curdef}). Close to the energy minimum 
the magnetization current along the $z$ axis  is
\bem
j_z= {\partial E\over \partial k_z} = {M-\langle M_z\rangle\over \gamma}{d \omega \over dk_z}
\nonumber \\
\approx {M-\langle M_z\rangle\over \gamma}{d^2 \omega(k_0) \over dk_z^2}(k_z-k_0).
       \eem{}
At our definition of the current  it is  proportional to the group velocity $d \omega / dk_z$ of magnons \cite{Adv} and therefore vanishes in the ground state of the condensate both for the running and the standing wave.

\section{Discussion and conclusions} \label{DC}

The derived critical gradient is essentially lower than obtained by \citet{Pokr} who determined the critical supercurrent equating the kinetic energy to the high Zeeman energy. Our analysis demonstrates that the Zeeman energy does not affect the stability condition at all. The magnetostatic term (\ref{mse}) stabilizing supercurrents plays the same role as easy-plane anisotropy in easy-plane magnets, but the former is of dynamical origin and  much smaller than the latter being proportional to the wave intensity (density of condensed magnons).

A byproduct of our analysis was revision of the widely accepted spin-wave spectrum in YIG films, which took into account 
 proper  magnetostatic and exchange boundary conditions on film surfaces. This influenced estimations of non-linear corrections to spin waves crucial for metastability of the magnon condensate with and without spin supercurrents.

Let us make some numerical estimations. According to \citet{Dem07} the magnon density can reach $10^{18}$ cm$^{-3}$. Assuming that 10 \% of magnons are in the coherent state, this corresponds  to rather small ratio $(M-\langle M_z \rangle)/M \sim 0.32\times 10^{-4}$.  Then  \eq{cr} yields for $k_0 =5.5~10^4$ cm$^{-1}$ and $d=5~10^{-4}$ cm the critical velocity $v_{cr}$ about 3.6 m/sec (instead of 420 m/sec found by \citet{Pokr}). 
  
 In the light of the presented analysis let us discuss the report by \citet{spinY}  on detection of spin supercurrents in observation of a decaying magnon condensate prepared in a YIG magnetic film by magnon pumping. The major problem with this claim is small  total phase variation along streamlines of the supposed current realized in the experiment. Bozhko {\em et al.} applied a temperature gradient to the magnon BEC cloud, which led to  a difference $\delta \omega$ of the  frequency of magnetization precession (phase rotation velocity) across the condensate cloud.   This produced a total phase variation $\delta \varphi =\delta \omega t$ across the BEC cloud growing linearly with time  $t$ and generating spin currents.  For the maximal $\delta \omega = 2\pi \times 550$ rad/sec and the maximal life time $t =0.5~\mu$sec of the condensate in the experiment of \citet{spinY} (see their Fig.~5) one can conclude that the total  phase variation $\delta \varphi$ never exceeded about 1/3 of the full $2\pi$ rotation.  As discussed in introduction, only currents  with large number of full $2\pi$ rotations along streamlines deserve the title of ``supercurrent'' manifesting spin superfluidity. 
 
 One might consider it as a purely semantic issue. But calling any current $\propto\nabla\varphi$   supercurrent demonstrating spin superfluidity would reduce spin superfluidity to a trivial ubiquitous phenomenon. 
Currents produced by such small phase variations cannot relax via phase slips and are trivially stable. They 
emerge in any spin wave or domain wall. Any inhomogeneity produces  them, and they  must present in the experiment of  \citet{spinY} but in contrast to authors' claim they  have nothing to do with the macroscopic phenomenon of superfluidity.  

In summary, spin superfluidity in YIG films is possible in principle, although the recent report on its experimental observation \cite{spinY} is not founded. Metastability of spin supercurrents in this material is provided by energetic barriers not of topological but of dynamic origin, which depend on intensity of a nonlinear spin wave describing the coherent magnon condensate. 

It is worth noting that at growing magnetic field in YIG films the orientational phase transition takes place from the state with the total and sublattice magnetizations  along the magnetic field to the state, in which magnetizations deviate from the magnetic field direction and have large components in the plane normal to the magnetic field.\cite{Melk} This is a state with easy-plane anisotropy, for which spin superfluidity have been predicted. But this requires  magnetic fields  $\sim 10^5$~G, which are orders of magnitude larger than fields nowadays used in experiments on magnon condensation.

\begin{acknowledgments}
The author thanks S.O. Demokritov, V.S. L'vov, V.L. Pokrovsky, and A.A. Serga for useful discussions.
\end{acknowledgments}


%

\end{document}